\documentclass[journal]{IEEEtran}
\usepackage{amsmath, amsthm, amssymb, amsfonts}
\usepackage{mathtools}
\usepackage{bm}
\usepackage{graphicx}

\begin{document}
\title{An optimal real-time controller  for \\ vertical plasma stabilization}
%
%

\author{N. Cruz, 
        J.-M. Moret, 
        S. Coda,
        B.P. Duval,
        H.B. Le,
        A.P. Rodrigues,
        C.A.F. Varandas,
        C.M.B.A.  Correia 
        \\ and B. Gon\c{c}alves
        
\thanks{Manuscript received June 16, 2014. This work was supported by EURATOM and carried out within the framework of the European Fusion Development Agreement. IST activities also received financial support from "Funda\c{c}\~{a}o para a Ci\^{e}ncia e Tecnologia" through project Pest-OE/SADG/LA0010/2013. The views and opinions expressed herein do not necessarily reflect those of the European Commission. This work was partly supported by the Swiss National Science Foundation.}
\thanks{N. Cruz is with the Instituto de Plasmas e Fus\~ao Nuclear, Instituto Superior T\'ecnico, Universidade de Lisboa, P-1049-001 Lisboa, Portugal (Telephone: +351-239410108, e-mail: nunocruz@ipfn.ist.utl.pt).}%
\thanks{J.-M. Moret, S. Coda, B. P. Duval and  H. B. Le are with the Centre de Recherches en Physique des Plasmas, \'Ecole Polytechnique F\'ed\'erale de Lausanne (EPFL), CH-1015 Lausanne, Switzerland.}%
\thanks{A. P. Rodrigues, C. A. F.  Varandas and B.S. Gon\c{c}alves are with the Instituto de Plasmas e Fus\~ao Nuclear, Instituto Superior T\'ecnico, Universidade de Lisboa, P-1049-001 Lisboa, Portugal.}%
\thanks{C. M. B. A.  Correia is with the Centro de Instrumenta\c{c}\~ao, Departamento de F\'isica, Universidade de Coimbra, P-3004-516 Coimbra, Portugal}%
}

\maketitle

\begin{abstract}
Modern Tokamaks have evolved from the initial axisymmetric circular plasma shape to an elongated axisymmetric plasma shape that improves the energy confinement time and the triple product, which is a generally used figure of merit for the conditions needed for fusion reactor performance. However, the elongated plasma cross section introduces a vertical instability that demands a real-time feedback control loop to stabilize the plasma vertical position and velocity. At the Tokamak à Configuration Variable (TCV) in-vessel poloidal field coils driven by fast switching power supplies are used to stabilize highly elongated plasmas. TCV plasma experiments have used a PID algorithm based controller to correct the plasma vertical position. In late 2013 experiments a new optimal real-time controller was tested improving the stability of the plasma.

This contribution describes the new optimal real-time controller developed. The choice of the model that describes the plasma response to the actuators is discussed. The high order model that is initially implemented demands the application of a mathematical order reduction and the validation of the new reduced model. The lower order model is used to derive the time optimal control law. A new method for the construction of the switching curves of a bang-bang controller is presented that is based on the state-space trajectories that optimize the time to target of the system.

A closed loop controller simulation tool was developed to test different possible algorithms and the results were used to improve the controller parameters.

The final control algorithm and its implementation are described and preliminary experimental results are discussed.

\end{abstract}


\section{Introduction}

\IEEEPARstart{M}{odern} tokamak devices \cite{Wesson} are designed to accommodate  elongated cross-section plasmas \cite{Ariola}\cite{Ambrosino2005} to improve fusion performance. A vertically elongated plasma presents important advantages since it allows the creation of divertor plasmas, the increase of the plasma current and density limit as well as providing plasma stability. However, an elongated plasma is unstable due to the forces that pull the plasma column upward or downward. The result of these forces is a plasma configuration that tends to be pushed up or down depending on the initial displacement disturbance. For example,  a small displacement downwards results in the lower poloidal field coils pulling the plasma down, with increased strength as the plasma gets further from the equilibrium position. To compensate this instability,  feedback controllers have been designed to correct the vertical position displacement \cite{Sartori2010}\cite{Vyas1998}\cite{Walker2009}.

The design of vertical stabilization feedback controllers has been based in simple models, resulting in experimentally tuned Single Input Single Output (SISO) Proportional Integral and Derivative (PID) regulators. This procedure requires an in-depth experimental treatment that is time consuming and demands a big number of experimental discharges to obtain the necessary gains optimization. This paper presents an alternative method to design the vertical stabilization controller of a tokamak using a simple plasma model and the application of optimal control theory.

This paper is organized as follows: 
Section II presents the vertical observer developed to detect the plasma centroid vertical position and velocity in real-time; 
Section III briefly depicts the different methods that can be used to describe a tokamak plasma; 
Section IV describes the state-space plasma model that predicts the plasma response to the actuators 
and the model reduction performed to permit the application of the time optimal control theory that is presented in Section V; 
Section VI depicts the simulation tool that permits off line testing and parameter tunning of the controller; 
The controller results and future work is presented in Section VII.

\section{Vertical Plasma Position Observer}

The vertical position observer is a linear combination of the magnetic field measured using the magnetic diagnostics. A matrix containing the contribution weight of each magnetic probe is calculated before each plasma discharge, taking into account the planned plasma parameters such as shape and position. The contribution of each probe to the observer has in account the pre-planned plasma parameters, because the probes closer to the plasma are more efficient estimating its position and will be given more weight in the observer. 

A set of coefficients are calculated to define the observer from a finite element set of plasma current filaments, using Green's functions, thus it is possible to calculate the magnetic field produced in the probes. The matrix is built with the set of probes that are going to be used in the measurements and inverted to obtain the observer coefficients \cite{Hofman1998}\cite{RT2010_NCruz}.

The following equation relates the magnetic field measurements with the currents in the plasma:
\begin{equation}
 b_{m}-B_{mc}.I_{c}=B_{mx}.I_{x}	 
\end{equation}
with $b_m$ the vector of measured quantities in the magnetic probes, $B_{mc}$ the matrix of the Green's functions between the coils and the magnetic probes, $I_c$ the coil currents vector, $B_{mx}$ the matrix with the Green's functions transforming the current in the plasma filaments into the magnetic field measured by the probes and $I_x$ the currents in the toroidal filaments.

From the inversion of the equation, the currents can be obtained by: 
\begin{equation}
 I_x=A^{-1}(B_{mx}^T.b_m-B_{mx}^T.B_{mc}.I_c) 	
\label{eq:2}
\end{equation}
where $A=B_{mx}^T.B_{mx}$. Equation (\ref{eq:2}) gives the current in the toroidal filaments as a linear combination of the magnetic field measured by each probe (first term) with the correction of the coil current influence in the measurements. 

The observer is then given by:
\begin{equation}
 zI_p=(z_{x}'-z_{a}).I_x	
\end{equation}
with $z$ the vertical position of the plasma centre, $z_{x}$ the position of the filament of the plasma column and $z_{a}$ the reference position of the plasma axis.

The plasma velocity observer ($d(zI_p)/dt$) uses the same method and because the time derivative of $I_c$ has a slow variation compared to vertical position growth rate, the equations can be reduced to:
\begin{equation}
\frac{dI_x}{dt}=A^{-1}.B_{mx}^T.\frac{db_m}{dt} 
\end{equation}
\begin{equation}
\frac{d(zI_p)}{dt}=(z_{x}'-z_{a}).\frac{dI_x}{dt}   
\end{equation}

The coil currents correction is added to the reference signal, which makes the output error signal completely consistent.

\section{Plasma Description}

The modeling of a tokamak plasma demands complex mathematical 
calculation, in depth physical knowledge and computational power for 
numerical calculation during simulation phase. Different paths have been tried 
to accomplish this mission.

The simpler models consider the plasma as a filament or non-deformable 
matrix of conducting filaments. The more complex models include nonlinear codes, which permit the simulation of nonlinear behaviors such as large vertical position 
displacements. Some important plasma model and reconstruction codes include \cite{Garrido}:
\begin{itemize}
	\item \textit{PET} is a free boundary plasma equilibrium evolution code developed at the  Efremov Scientific Research Institute, St. Petersburg \cite{PET}.
	\item \textit{ASTRA} \cite{Astra} (Automated System for TRansport Analysis) is a code to solve a set of transport equations in toroidal geometry. This code is presently used to make transport simulations of tokamak and stellarator plasmas. The first version of ASTRA was implemented at the Kurchatov Institute in Moscow, but an international community continues to develop the code and new features are added to its functionality regularly.
	\item \textit{TSC} (Tokamak Simulation Code) was originally developed by S. C.  Jardin at Plasma Physics Laboratory, Princeton University for free boundary 2D transport \cite{Jardin_TSC}.
	\item \textit{EFIT} (Equilibrium FITting) is a code developed to perform magnetic and kinetic-magnetic analysis for Doublet-III, at General Atomics. EFIT takes the measurements from plasma diagnostics and calculates relevant plasma properties such as geometry, stored energy and current profiles. Although it is a very fast computational code, it lacks the accuracy of other more computational intensive algorithms \cite{EFIT}.
	\item \textit{FBT} (Free Boundary Tokamak) is a code originally developed by F. Hofmann at Centre de Recherches en Physique des Plasmas, \'Ecole Polytechnique F\'ed\'erale de Lausanne. FBT allows the computation of arbitrarily shaped tokamak equilibrium specially dedicated to highly shaped and elongated plasmas \cite{Hofmann_FBT}.
	\item \textit{PROTEUS} is a nonlinear tokamak simulation code that solves the Grad-Shafranov equation  by an iterative finite element method. This code is used to simulate the evolution of a tokamak plasma for a fixed plasma current \cite{PROTEUS}.
	\item \textit{CREATE-L} is a linearized plasma equilibrium response model in view of the current, position and shape control of plasmas in tokamaks \cite{CREATE-L}\cite{Vyas}. The origin of this code's name is the consortium where it was originally developed, the \textit{Consorzio di Ricerca per l' Energia e le Applicazioni Tecnologiche dell'Elettromagnetismo} (CREATE). 
	\item \textit{DINA} is a tokamak plasma axisymmetric, time-dependent, resistive
MHD simulation code and a free boundary equilibrium solver developed at the RRC Kurchatov and TRINITI institutes in Moscow  \cite{DINA}. 
  \item \textit{RZIP} is a rigid plasma model that predicts the plasma current, as well as the radial and the vertical plasma positions, used at Centre de Recherches en Physique des Plasmas, Lausanne \cite{Coutlis1999}.  
\end{itemize}

Some of these codes are accurate for plasma simulation and reconstruction but due to its complex structure are not suitable for controller design. This action is based on simpler linear models that ensure the stability, robustness and performance of the controller, provided that the states are not too far from equilibrium. Controllers are thus usually 
designed based on the linear model of the flat top phase, achieving good 
performance through the whole discharge due to its robustness.

Linear models for control design purposes use the electrical circuit
equations to calculate the time evolution of the plasma current. Two
such models are the CREATE-L and the RZIP models. CREATE-L considers the plasma deformation through the calculation 
of the plasma current distribution equilibrium. On the other hand, RZIP is an enhanced non 
deformable model that may vary its vertical and radial position, as well
as its total plasma current. The RZIP model is presented in the next section, to be used for the design of the new plasma stabilization controller.

\section{Plasma Model for Control}

\subsection{RZIP Plasma Model }

The use of the RZIP plasma model aims at finding the transfer function between the currents in the poloidal field coils, internal to the TCV structure close to the plasma, and the vertical plasma displacement \cite{Lazarus1990}\cite{Lister1990}\cite{Lister1996}\cite{Hofmann1997}.

The RZIP model gets its name from the simplifications assumed to build the circuit equations, with the following characteristics: (i) the current has constant distribution, rigid model, as the plasma shape is assumed not to change; (ii) the center of the vertical position can change: plasma is free to move vertically; (iii) the center of the radial position can change: plasma is free to move radially; (iv) the integral of the plasma filaments current can change: the total plasma current is free to change.

The model design simplifications give important advantages over more complex plasma models, maintaining an overall accuracy: (i) A simple model that is easier to implement; (ii) No need to calculate the complete plasma equilibrium; (iii) More explicit model to the quantities that define plasma response to the control variables (a better control model).	

The model is derived from (i) the equilibrium equation of the vertical forces in the plasma; (ii) the equilibrium equation of the radial forces in the plasma and (iii) the plasma current circuit equations \cite{Coutlis1999}, resulting in an equation that includes the output voltages of the power supplies, the currents in the control coils and the plasma position and current:

\begin{equation}
\textbf{M}sx + \bm{\Omega} x = u
\label{eq:model1}
\end{equation}
where $s$ is the Laplace variable, and the matrices are given by: 

\begin{equation}
\textbf{M} = \left( 
\begin{array}{cccc}
\textbf{M}_c  &  (M'_z)^{T}  &    (M'_R)^{T}           &               (M_p)^{T}             \\
M'_z               & \alpha              & 0          & 0          \\	
M'_R               &  0                  & M_{33}     & M_{34}           \\
M_p                &  0                  & M_{43}     & L_{p0}           \\
\end{array}
\right)
 \text{ ; }
\end{equation} 

\begin{equation}
x = \left( 
\begin{array}{c}
\bm{I}_c \\
zI_{p0}    \\
RI_{p0} \\
I_p \\
\end{array}
\right)
 \text{ ; }
\end{equation} 

\begin{equation}
u = \left( 
\begin{array}{c}
\bm{V_c} \\
0    \\
-\mu_0I_{p0}^2 s \Gamma \\
0 \\
\end{array}
\right)
 \text{ ; }
\end{equation} 

\begin{equation}
\bm{ \Omega } = \left( 
\begin{array}{cccc}
\Omega_c    &  0                  & 0          & 0           \\
 0                 &  0                  & 0          & 0           \\
0                  &  0                  & 0          & 0           \\
0                  &  0                  & \Omega'_p   & \Omega_p           \\
\end{array}
\right)
\end{equation} 

The values in the Mutual matrix are given by:

\begin{equation}
M_{33}= \left[ \frac{\mu_0}{2} \frac{\partial\Gamma}{\partial R} +  \frac{2\pi}{I_{p0}}
\left(B_{z0}+R_0 \frac{\partial B_z}{\partial R}\right) \right]
\end{equation} 

\begin{equation}
M_{34}= \mu_0 \Gamma_0 + \frac{2\pi R_0 B_{z0}} {I_{p0}}
\end{equation} 

\begin{equation}
M_{43}= \mu_0 (1+f_0) + \frac{2\pi R_0 B_{z0}} {I_{p0}}
\end{equation} 


The equation was also derived to a state-space model. A function reads the plasma equilibrium details from the TCV database based on the discharge number and time, as well as the tokamak structure parameters used to build the state-space model for specific plasma elongations. 

\subsection{Step Response to a Voltage Change on the Fast Coil}

From the complete RZIP model described with some state variables that can be neglected in the vertical stabilization problem, the model was simplified aiming at calculating the transfer function from the current on the internal FPS coils to the plasma vertical position. This is the mathematical method that describes the influence of the currents in the fast coils in the plasma vertical position.
 
The simplification of the full plasma model for the particular case of the plasma vertical stability using the in-vessel fast coils presents a difficulty from the fact that the multiple input multiple output (MIMO) system that is obtained from the plasma model must be diagonalized to obtain a single input single output (SISO) system, independent from the remain system inputs and outputs. This is not always possible and some constraints must be analyzed to make them independent.

This simplification is possible for the present case because the vertical stabilization operates in a different time scale from the other plasma control variables such as position, shape or current. Moreover the vertical position that is also controlled by the poloidal coils outside the vessel can be considered independent of the internal poloidal coils, because of the same reason. While the poloidal coils outside the vessel control the slow vertical displacement of the plasma, the in-vessel coils act on a much faster time-scale, reacting to fast plasma disturbances.

The state space system is diagonalized to obtain the independent influence of the coil currents over the plasma vertical position. Then the equation of the fast coil is taken by neglecting the influence of the other coils. This is possible taking into account the referenced different time scale of the actuation of the coils. 
The typical way to address the vertical stabilization problem is to independently control the vertical plasma position from the plasma current and shape controllers \cite{Ariola}, which are designed on the basis that the system is vertically stable due to the controller already implemented. This double loop arrangement simplifies the design of the controllers, based on the assumption and later confirmation that the controllers act on different time scales. Different frequencies in the controllers permit the treatment of some parameters as disturbances to the next stage of the global controller. 

\subsection{Model Reduction and Validation}

In control engineering, the best model is not always the most accurate, but the one that permits the construction of a robust stable controller, according to the necessary performance and specifications.

For the purpose of applying optimal control theory to the plasma model obtained a model reduction was necessary to permit the mathematical treatment presented in the next section. The transfer function that was obtained is of $52^{nd}$ order, while optimal control theory is usually applied to systems with second or third order at most. This led to the application of model reduction techniques.

Model reduction techniques are a powerful tool that uses methods based on the idea of projecting the state space to a much lower dimension, obtaining a reduced system that may  be solved more efficiently. 
For control design purposes, it is possible to approximate the model with another model of reduced order that preserves the original transfer function as much as possible. 

\begin{figure}
	\centering
	\includegraphics[scale=0.60]{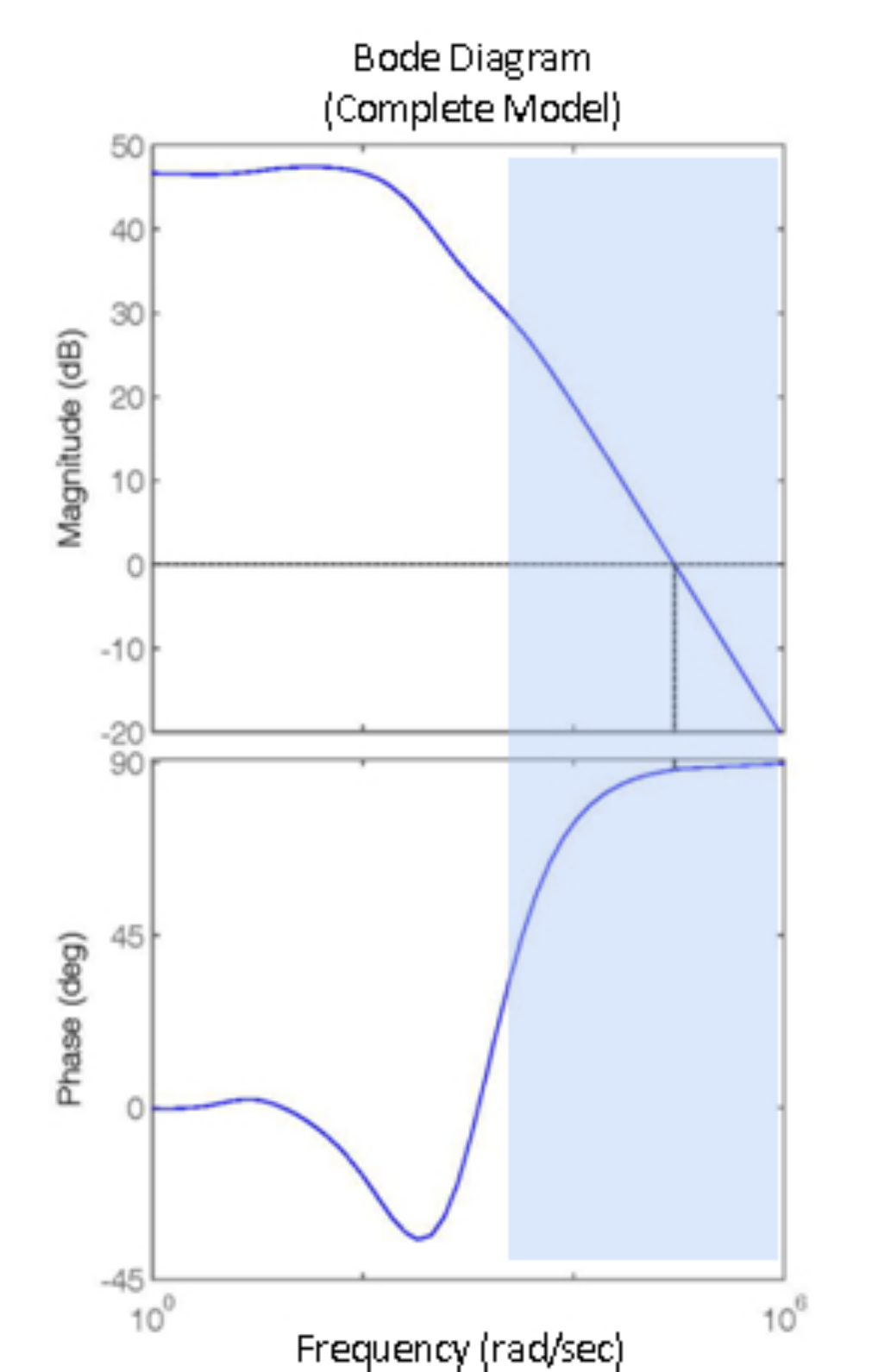}
	\caption{Bode diagram of the complete model transfer function  }
	\label{fig:BodeDigramCompleteModel}
\end{figure}

\begin{figure}
	\centering
	\includegraphics[scale=0.60]{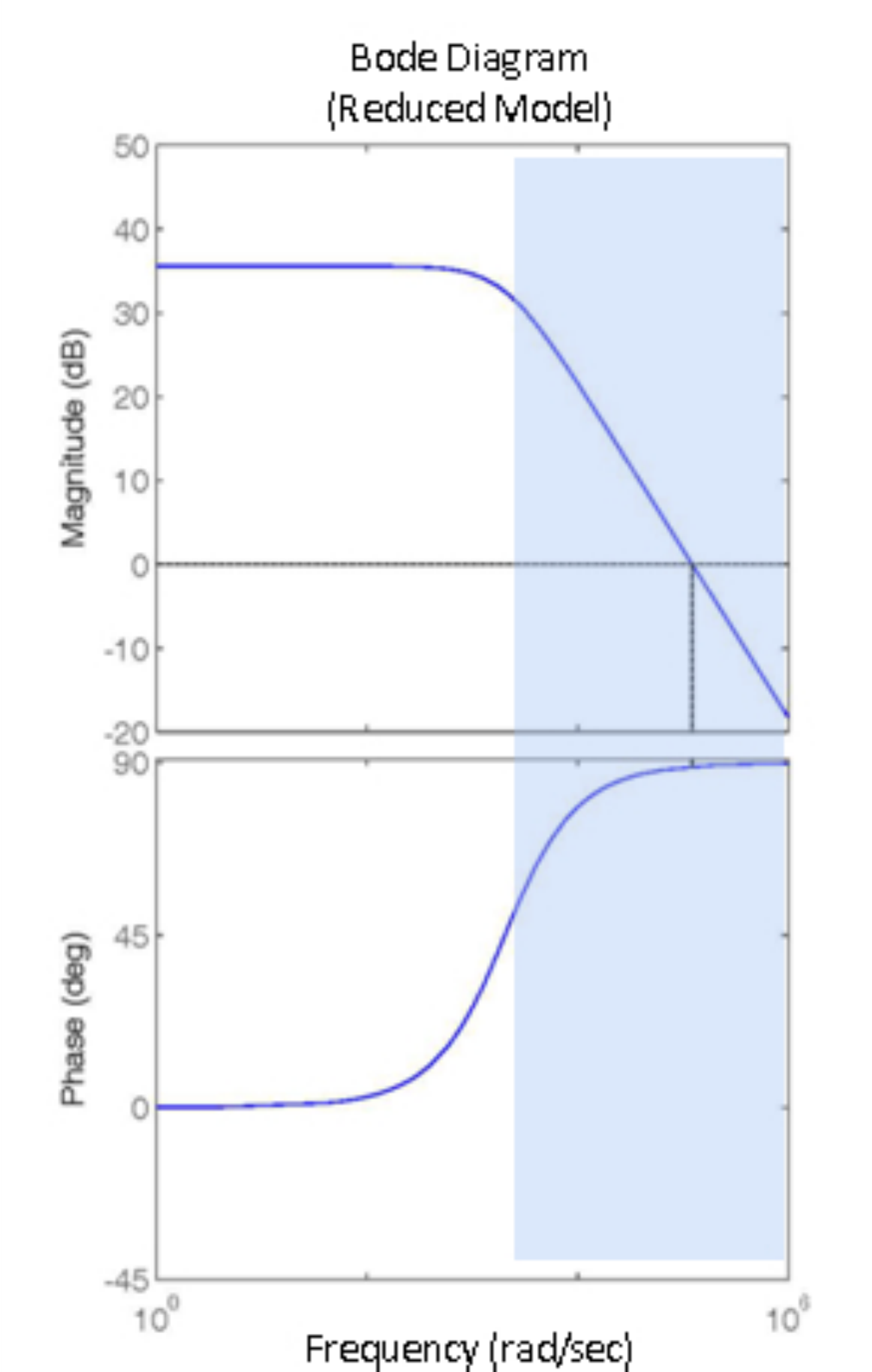}
	\caption{Bode diagram of the reduced model transfer function }
	\label{fig:BodeDigramReducedModel}
\end{figure}

The method of balanced realization was applied to reduce the transfer function \cite{Laub}, by eliminating the states with small $\sigma_i$, i.e. with small influence in the behavior of the transfer function. This method permits the model reduction to a second order transfer function with difference results that could not be detected by the plot of the step response of both transfer functions. In the bode diagram plot of both models (Figures \ref{fig:BodeDigramCompleteModel} and \ref{fig:BodeDigramReducedModel}) differences were detected but only on slower frequencies that are not relevant for plasma vertical stabilization. The blue shadow areas in the figures show the agreement between both models for the frequencies of interest.

\section{Optimal Control}

\subsection{System Definition}
This section describes the application of the  time optimal control to the second order model to obtain a control law, the switching time and the final time of the bang-bang controller \cite{Athans}\cite{Vakilzadeh1978}\cite{Vakilzadeh1982}\cite{Shen}.

The second order transfer-function that describes the plasma model has the form:
\begin{equation}
\dfrac{X_{s}(s)}{U(s)}=\frac{n_{1}s+n_{2}}{s^{2}+d_{1}s+d_{2}}
\end{equation}

This transfer function represents the following controllable state space model:

\begin{equation}
\dot{X}=AX+Bu
\end{equation}

with 
$X=\left[\begin{array}{c} x_{1}\\x_{2}\end{array}\right]$, 
$A=\left[\begin{array}{cc}0 & 1\\-d_{2} & -d_{1}\end{array}\right]$, 
$B=\left[\begin{array}{c}b_{1}\\b_{2}\end{array}\right],$
where 
\begin{equation}x_{1}=x_{s}\end{equation}
and 
\begin{equation}x_{2}=x_{1}+b_{1}u \end{equation}
are the system variables when $b_{i}$ is given by 

\begin{equation}
\left[\begin{array}{c}0\\n_{1}\\n_{2}
\end{array}\right]=\left[\begin{array}{ccc}
1 & 0 & 0\\
d_{1} & 1 & 0\\
d_{2} & d_{1} & 1
\end{array}\right]\left[\begin{array}{c}
b_{0}\\
b_{1}\\
b_{2}
\end{array}\right]
\end{equation}

The eigenvalues of A are thus, given by 

\begin{equation}
\lambda_{1}=-\frac{n_{1}}{2}+i\sqrt{4n_{2}-n_{1}^{2}}
\end{equation}

\begin{equation}
\lambda_{2}=-\frac{n_{1}}{2}-i\sqrt{4n_{2}-n_{1}^{2}}
\end{equation}

and the eigenvectors are given by

\begin{equation}
P=\left[\begin{array}{cc}
1 & 1\\
\lambda_{1} & \lambda_{2}
\end{array}\right]
\end{equation}

Having defined the system model and given the initial system state $X_{0}$, 
the aim is finding the control law  and parameters that take the system from 
the initial  state $X_{0}$ to a target state  $X_{1}$, minimizing the time to target. 

\subsection{Control Law}

The problem of finding the control law that drives the plasma position from an initial position $X_{0}$ to a final position $X_{1}$ in the minimum amount of time, is easier with the definition of a new system state $X_{N}$ and the redefinition of the state system equations. In this state system the set point becomes the origin, thus simplifying the problem:

\begin{equation}
X_{N}=X-X_{1} 
\end{equation}

\begin{equation}
\dot{X}_{N}=AX_{N}+Bu+AX_{1} 
\end{equation}

\begin{equation}
X_{N0}=X_{0}-X_{1} 
\end{equation}

Using Pontryagin's Minimum Principle (PMP), the aim is to minimize the cost function given by the time to achieve the set point:

\begin{equation}
J=\int_{0}^{t_{f}}dt
\end{equation}

According to PMP the control must minimize the optimal control theory Hamiltonian of the system that is given by:

\begin{equation}
H=1+\lambda^{T}(AX_{N}+Bu+AX_{1})
\end{equation}

where $\lambda$ is the state of the adjoint system, representing the system as a linear transformation using the vector space defined by the eigenvectors.

The combined system is thus given by:

\begin{equation}
\dot{X}_{N}=\frac{\partial H}{\partial \lambda}=AX_{N}+Bu+AX_{1} 
\end{equation}

\begin{equation}
\dot{\lambda}=-\frac{\partial H}{\partial X_{N}}=-A^{T}\lambda 
\end{equation}

Since $H(t)=0$ for all the time, we can conclude:

\begin{equation}
H(t_{f})=0 \underset{\dot{}}{\Rightarrow} 1+\lambda(t_{f})^{T}(AX_{N}(t_{f})+Bu(t_{f})+AX_{1})=0
\end{equation}

Moreover, using the information that $X_{N}(t_{f})=0$ because the target state is the origin, the previous equation may be simplified to
\begin{equation}
 1+\lambda(t_{f})^{T}(Bu(t_{f})+AX_{1})=0
\end{equation}

The minimization of this Hamiltonian yields the optimal time control law
\begin{equation}
\begin{array}{rcl}
 \lambda^{T}B>0 \underset{\dot{}}{\Rightarrow} u=u_{-} \\ 
 \lambda^{T}B<0 \underset{\dot{}}{\Rightarrow} u=u_{+}
\end{array}
\end{equation}

The bang-bang control law is complete with an arbitrary value of $u$ for $\lambda^{T}B=0$, which might also be a dead zone where no control is applied to avoid unnecessary switching due to hysteresis.

\subsection{Predictive Control and Construction of Switching Curves}

This subsection presents the method to predict the action ahead, preventing situations when the observer becomes temporarily unavailable, for example in the presence of Edge Localized Modes (ELMs). By the use of this method, it is possible to keep the system stable, by predicting the control action needed, provided the time the observer is not available is shorter than the final control time calculated and no other major unpredicted disturbance affects the system.

This method is based on the a-priori calculation of the switching time and final time for the optimal time control law of the system. This application uses some of the deduction and results already presented \cite{Shen}, but a new simpler and more generic algorithm was developed. The idea is to find the position where the following two paths cross each other. From the initial system state is applied the maximum control possible in the direction of the set point tracing this path. Also trace the path from the set point applying the opposite control backward in time. The state-space point where both paths cross is the place where the controller should switch. 
Based on the idea presented the following 5 step fully computational algorithm was developed and implemented:

\begin{description}
 \item [Step 1] Define what path control ($u_{max}$/$u_{min}$) should the system travel first in the direction of the set point, based on the initial system state.
 \item [Step 2] Build the trace of the path that the system travels from initial position, when the maximum/minimum control is applied $u_{max}$/$u_{min}$. The path is an array with system state and time information.
 \item [Step 3] Build the back trace in time that the system travels, when the maximum/minimum control is applied. This path includes a negative time array that counts the time from $t_f$ backward.
 \item [Step 4] Calculate the intersection of both paths, leading to the calculation of the desired values. The system state intersection time in the first array gives the switching time $t_s$, that can be added to the time in the system state of the second array to give the final time $t_f$.
 \item [Step 5] Repeat the same procedure to a different initial system state to find a matrix of initial system states versus switching and final times.
\end{description}

\section{Controller Simulations}

\subsection{Fast Power Supply Transfer Function}

To drive the current in the fast controller coils, a fast power supply is used. Because the coils are connected in series, although with opposite current directions, only one power supply is needed to drive the coils.

A model of the behavior of the power supply was built to be introduced in the controller simulations. The information from the fast power supply (FPS) and  internal poloidal coils \cite{Favre} was used to build the transfer function that permits the calculation of the current in the coils given the control signal sent to the power supply. The transfer function with gain $k$ and integration time $\tau$ is given by the equation:
\begin{equation}
G(s)=\frac{k}{\tau s+1}
\end{equation}

\begin{figure}
	\centering
	\includegraphics[scale=0.43]{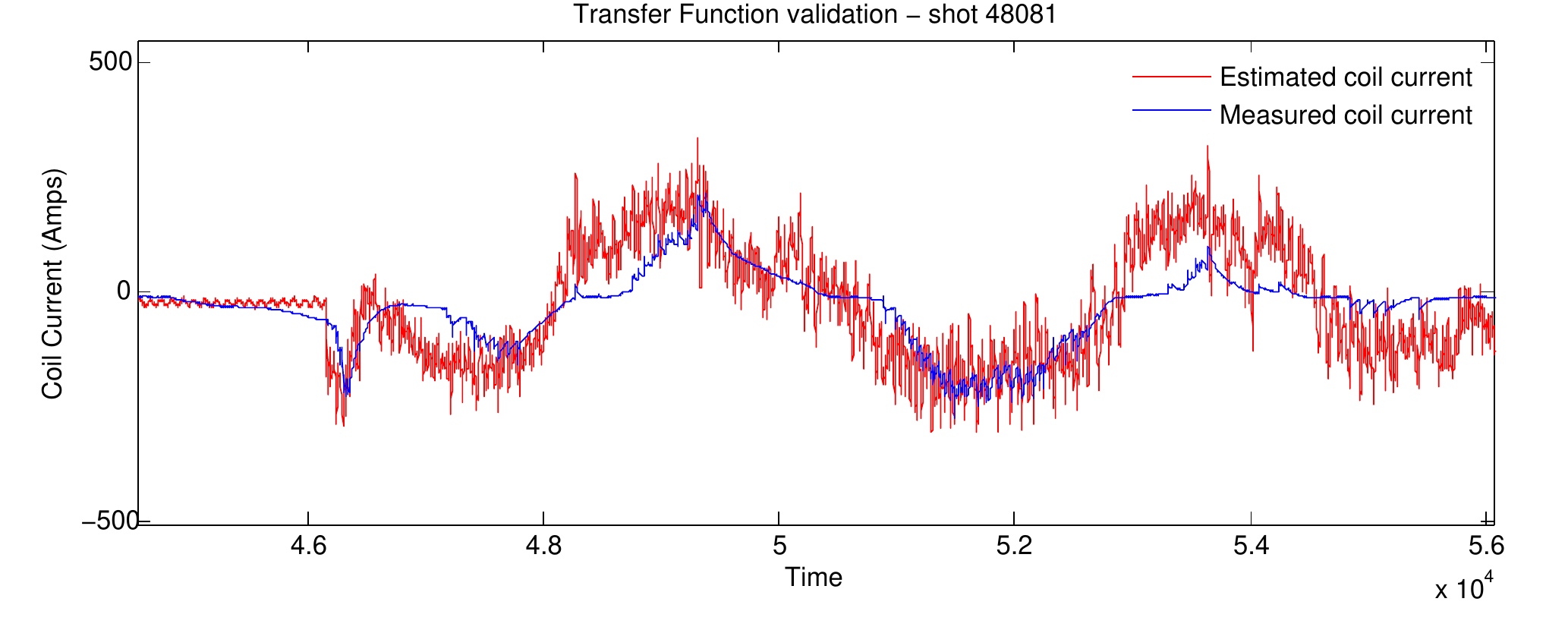}
	\caption{Estimated coil current using the transfer function over the FPS control signal}
	\label{fig:FPS_TF_Validation}
\end{figure}

Figure \ref{fig:FPS_TF_Validation} depicts the estimated coil current following the measured coil current using the FPS Transfer Function (TF).

\subsection{Simulator Tool}

\begin{figure*}
	\centering
	\includegraphics[scale=0.92]{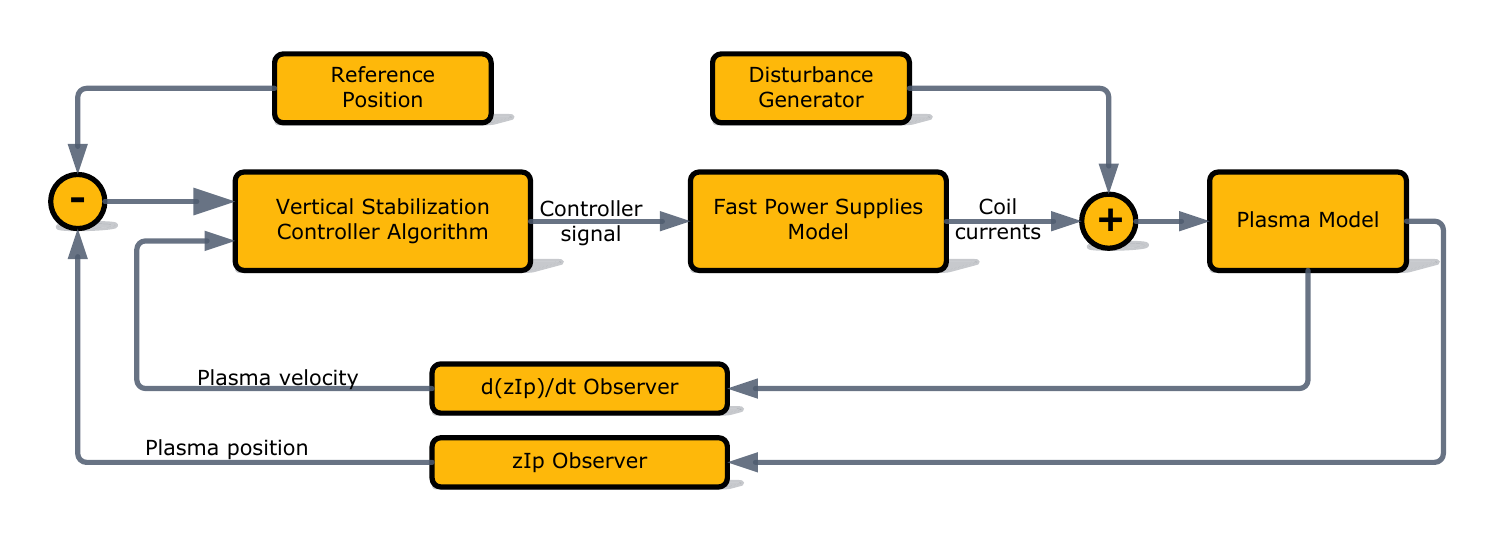}
	\caption{Block diagram of the simulation tool to analyze controller performance before implementation in real plasma discharges}
	\label{fig:SimulationTool2}
\end{figure*}

The plasma model was used to build a system simulation tool using Matlab Simulink  \cite{Simulink} (figure \ref{fig:SimulationTool2}). 
The simulator was implemented to test the different controllers and permit the fine tuning of any parameters before the use in real discharges in the tokamak. 


The plasma model includes the transfer function between the currents in the internal poloidal field coils and the plasma position, but lacks the transfer function of the fast power supplies that were also taken into account using a different simulation block. The stabilization controller has two inputs: the plasma velocity and plasma position error. From the inputs this block builds the controller signal to be sent to the fast power supplies. A disturbance generator is used to simulate unpredictable influences in the plasma. The complete plasma model is used for the simulation, for accuracy, because there is no need to use a reduced model except for the fact of faster computational simulations. Finally, the plasma model outputs the plasma position and a derivative block is used to simulate the plasma velocity measurements.

This Matlab Simulink model was used to obtain preliminary results.


\subsection{Controller Simulations}

The controller algorithm was tested and tuned based on simulation analysis. The decision for the best controller based on these analysis, resulted in a controller that adapts its force to the initial velocity  detected. 

A true bang-bang controller that always applies the maximum restore signal would exhibit a big oscillation in the plasma position. On the opposite side, a bang-bang controller that was limited to use a small control signal avoiding to exhibit oscillations, would be limited to the control of small perturbations. Thus, a weighted bang-bang controller that increases its restore signal according to the initial plasma velocity demonstrated to be much more efficient, resulting in a more stable controller. 

\begin{figure}
	\centering
	\includegraphics[scale=0.28]{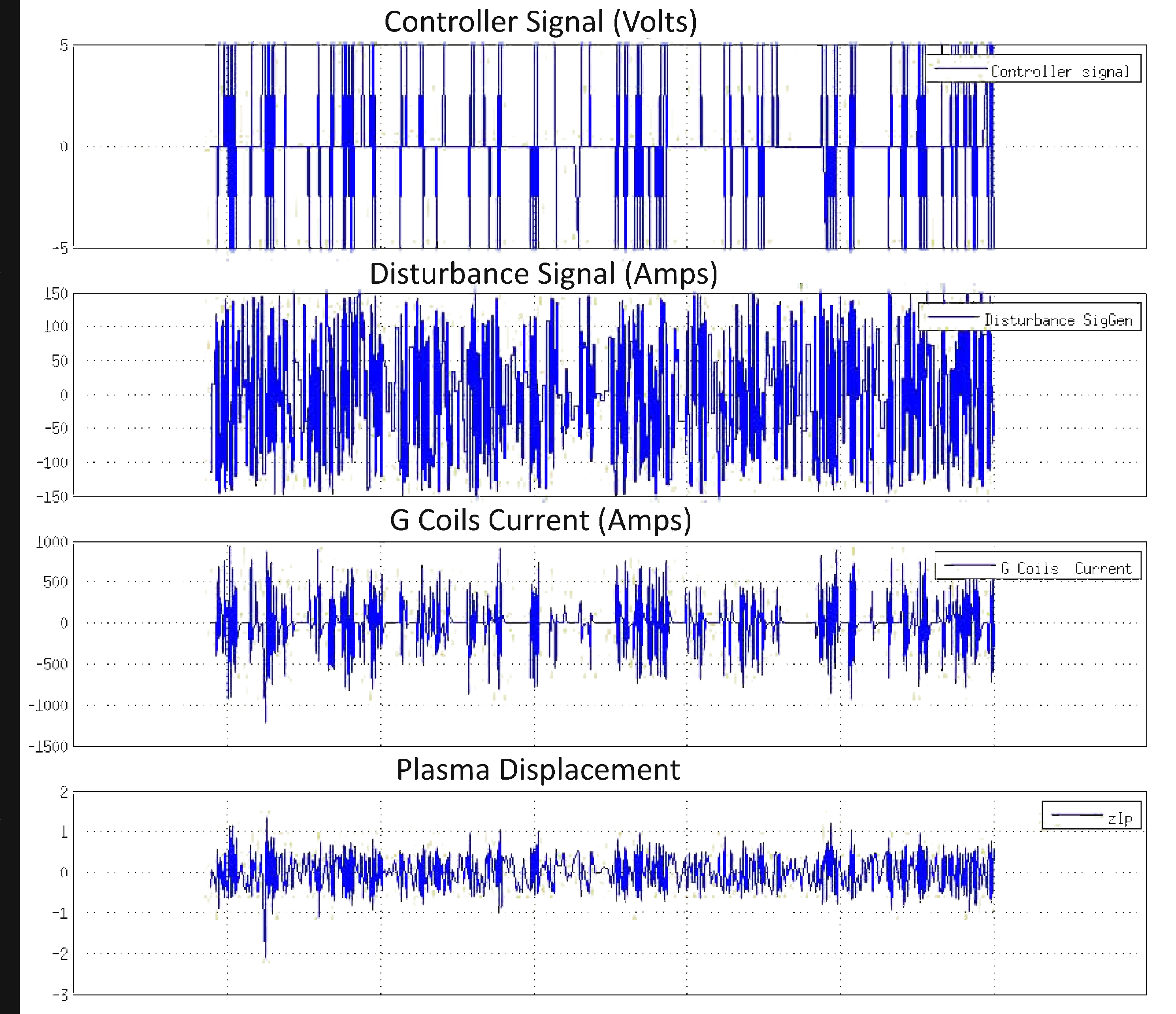}
	\caption{Simulation results of the bang-bang controller.}
	\label{fig:Simulation_2}
\end{figure}

\begin{figure}
	\centering
	\includegraphics[scale=0.28]{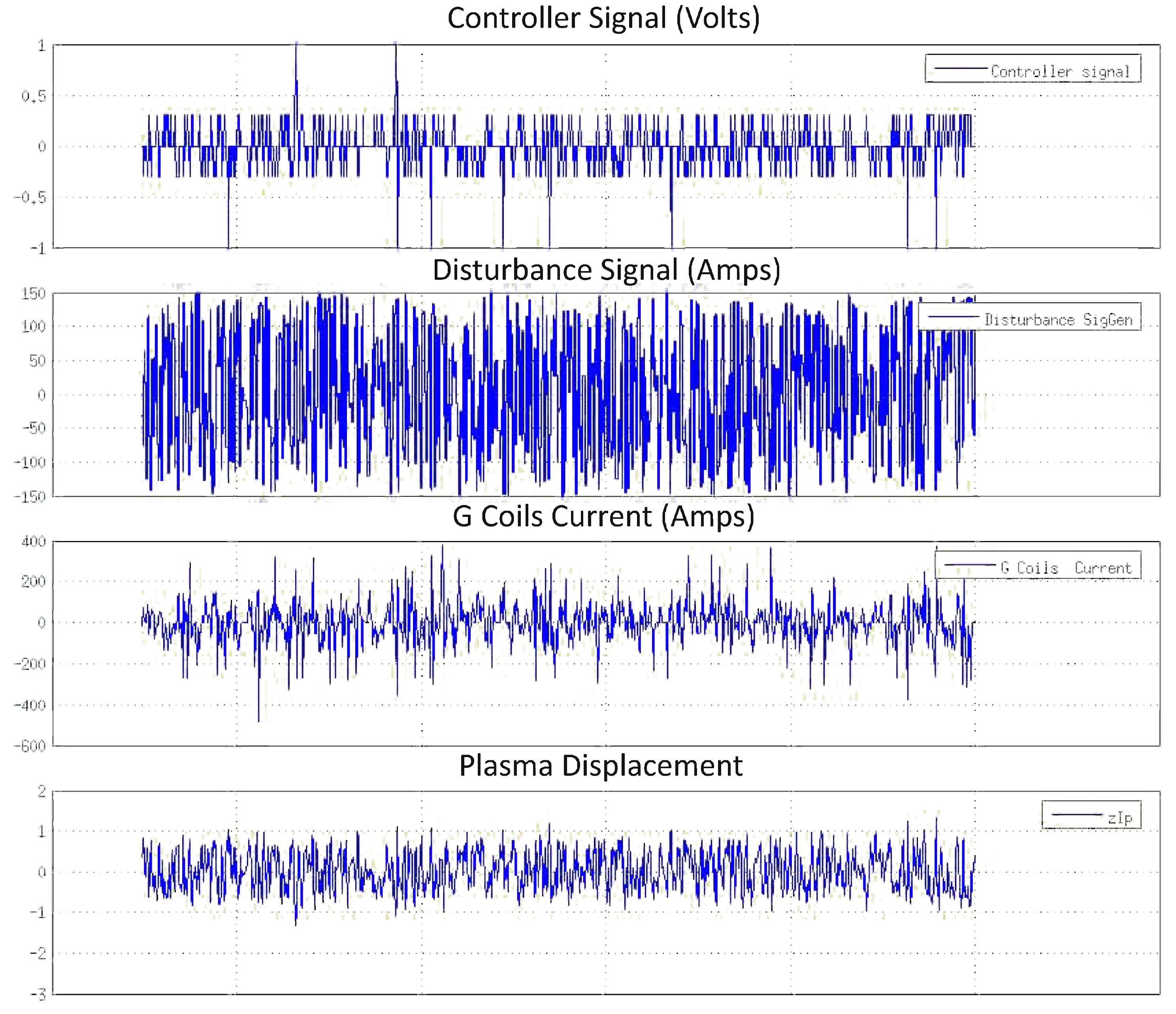}
	\caption{Simulation results of the variable bang-bang controller.}
	\label{fig:Simulation_1}
\end{figure}

Figures \ref{fig:Simulation_2} and \ref{fig:Simulation_1} support the use of a weighted bang-bang controller.

In these simulations it is possible to see a bang-bang controller with maximum possible strength that was tested against a high level of disturbances (fig. \ref{fig:Simulation_2}) with the plasma position under good control. However, using a variable bang-bang controller that changes state according to the distance of the plasma to the set point (fig. \ref{fig:Simulation_1}), also on the presence of big disturbances, the coil currents needed to stabilize the plasma are lower, as well as the plasma position error. The analysis of further simulations show that big disturbances can be controlled using a high control signal for higher displacements and smaller control signal for smaller displacements. 

\begin{figure*}
	\centering
	\includegraphics[scale=0.80]{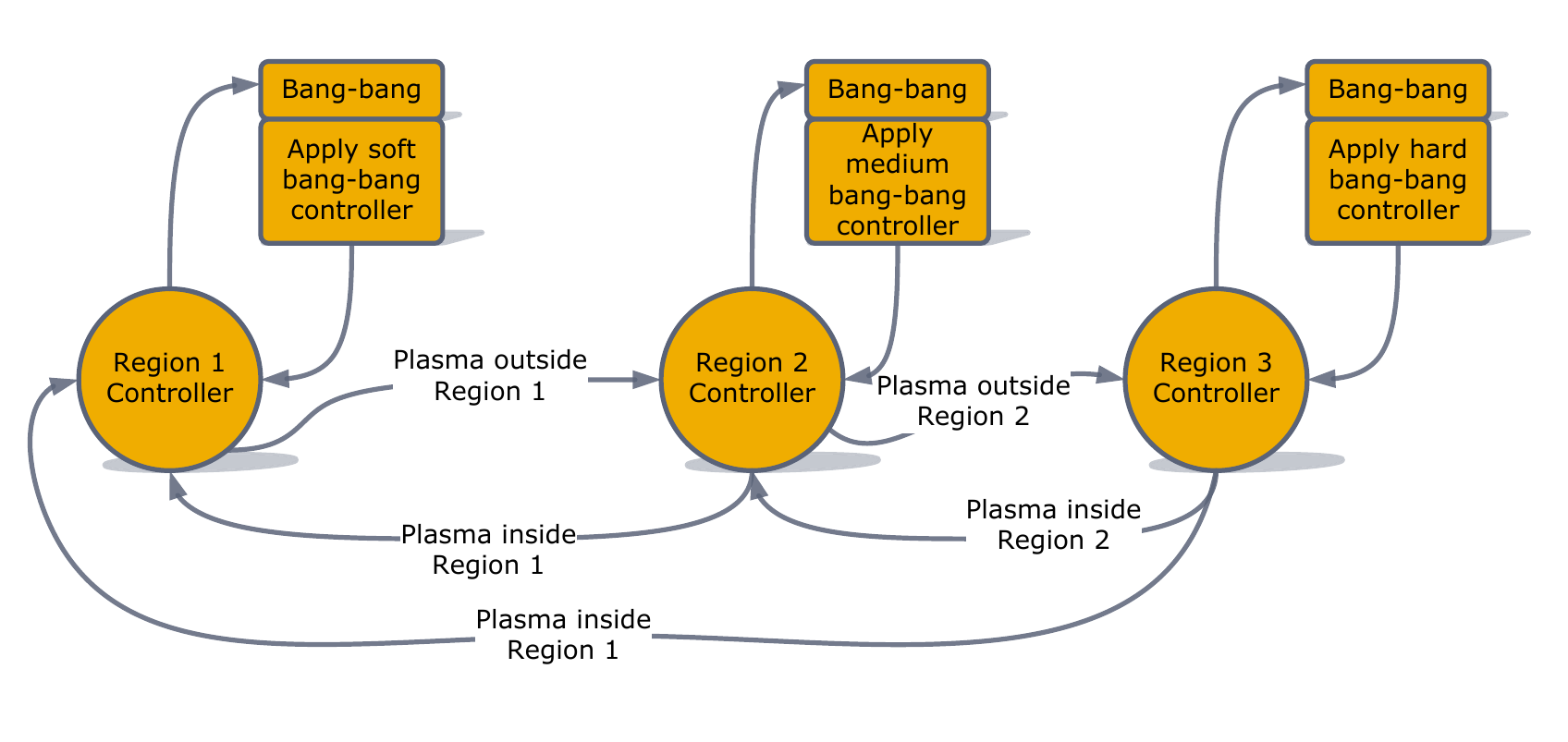}
	\caption{Diagram of the controller state machine}
	\label{fig:Controller_state_diagram}
\end{figure*}

Figure \ref{fig:Controller_state_diagram} represents a diagram with the controller state-machine. The  controller is a weighted bang-bang controller, that is similar to use an adaptive bang-bang controller that reconfigures based on system state position and velocity limits. This controller option improves stability by introducing a linear component to the classical nonlinear bang-bang controller.

\section{Controller Validation and Results}

\begin{figure}
	\centering
	\includegraphics[scale=0.38]{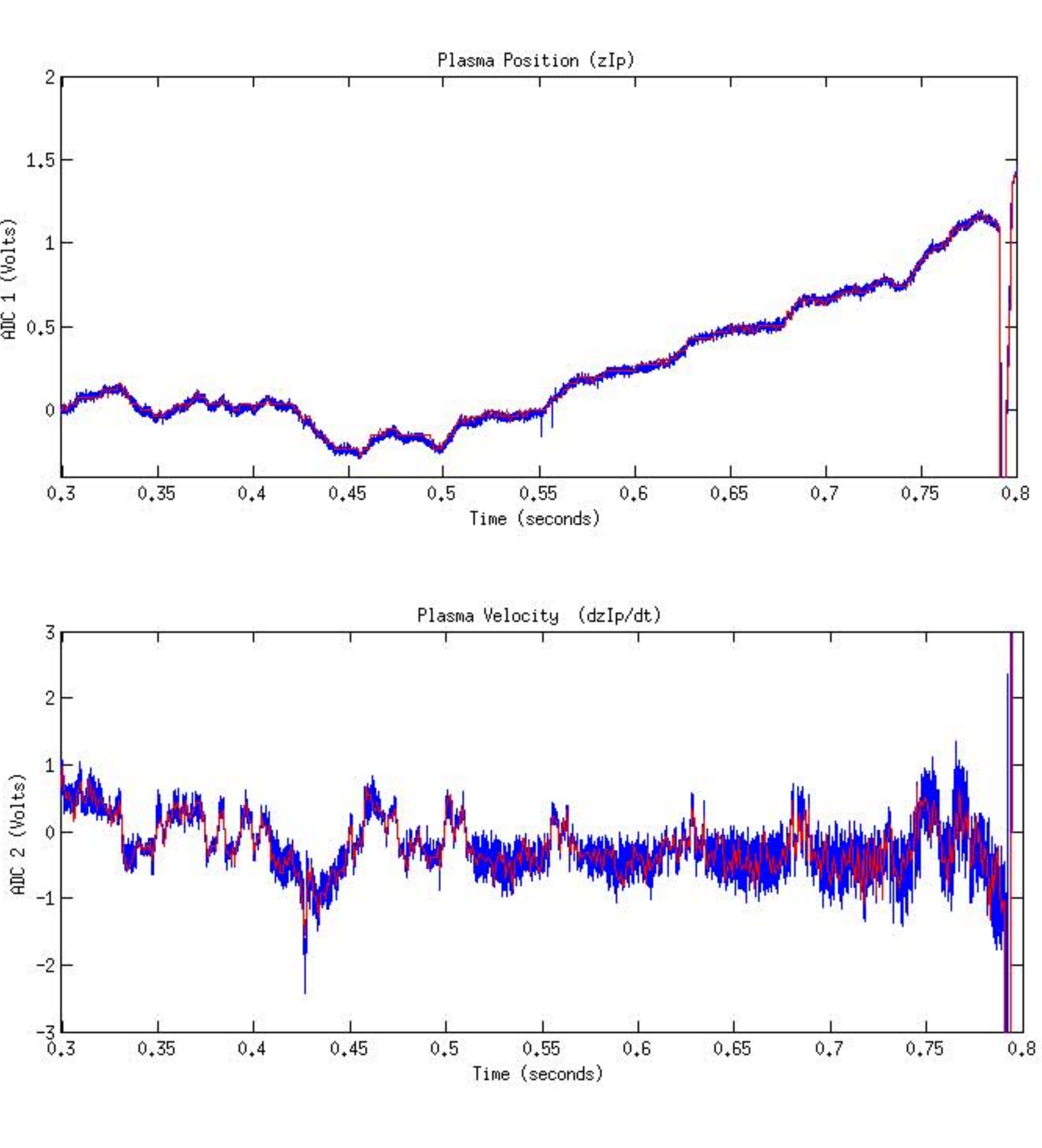}
	\caption{Plasma position and velocity for discharge 49564 using the new bang-bang controller}
	\label{fig:PlasmaPositionandVelocity49564}
\end{figure}

\begin{figure}
	\centering
	\includegraphics[scale=0.38]{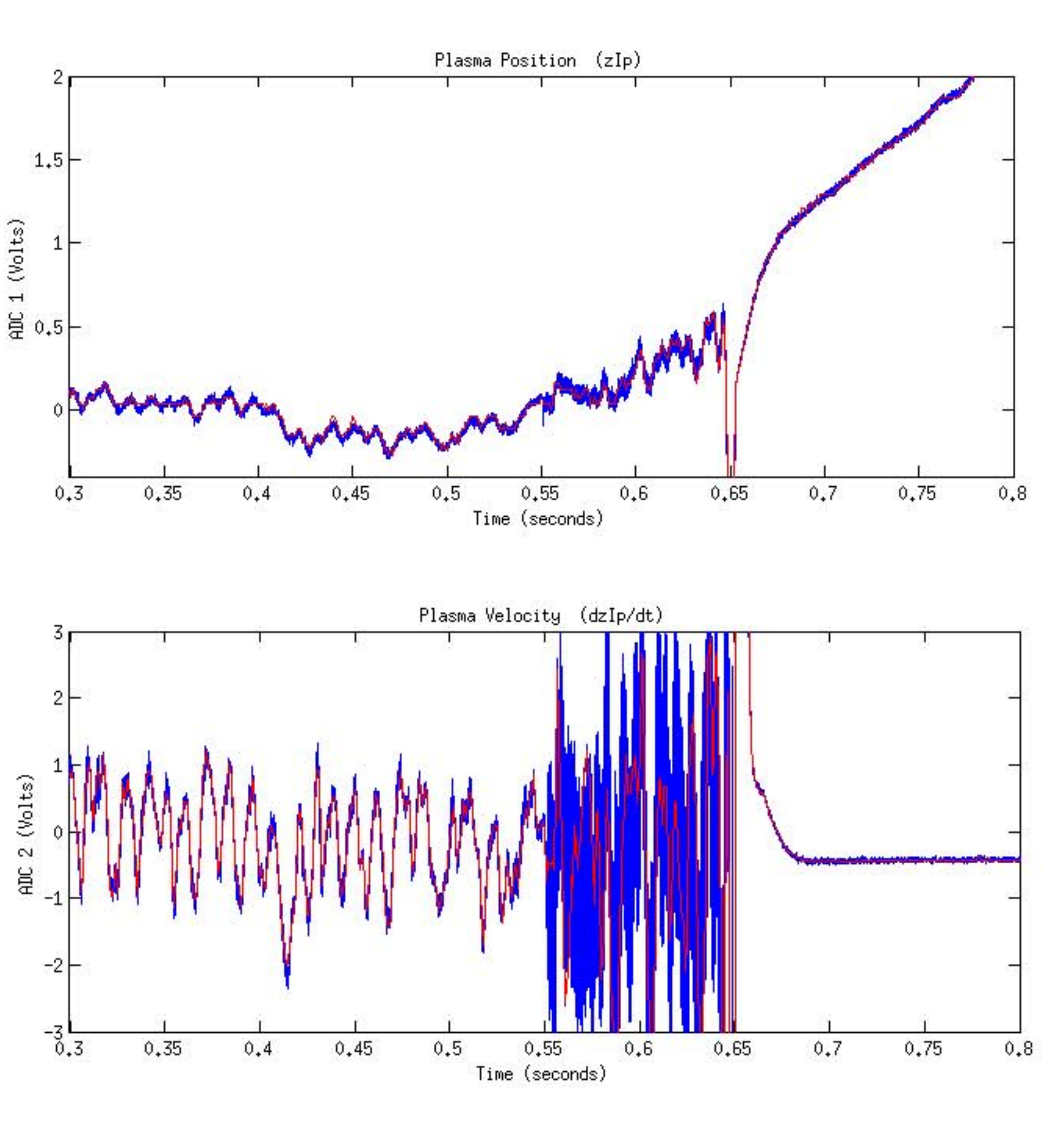}
	\caption{Plasma position and velocity  for reference discharge 49567 using the standard controller in the same plasma conditions as discharge 49564}
	\label{fig:PlasmaPositionandVelocity49567}
\end{figure}

The controller was implemented based on the simulation results and tested during plasma discharges at TCV, with improvement in the overall stability of the plasma. Figures \ref{fig:PlasmaPositionandVelocity49564} and \ref{fig:PlasmaPositionandVelocity49567} depict the stability improvement using the new controller. The plasma discharges were designed to test the limits of the controllers by increasing the plasma elongation from 0.5 seconds. 

The increased instability limit using the new controller can be confirmed by the improvement in discharge time for the same conditions. The current PID controller was not able to cope with the vertical instability finishing the discharge with a vertical disruption at approximately 0.65 s (0.15 s after starting the linear increase in plasma elongation). On the other hand the new bang-bang controller maintained the plasma discharge up to approximately 0.8 s (0.3 s after starting the linear increase in plasma elongation).

\begin{figure}
	\centering
	\includegraphics[scale=0.38]{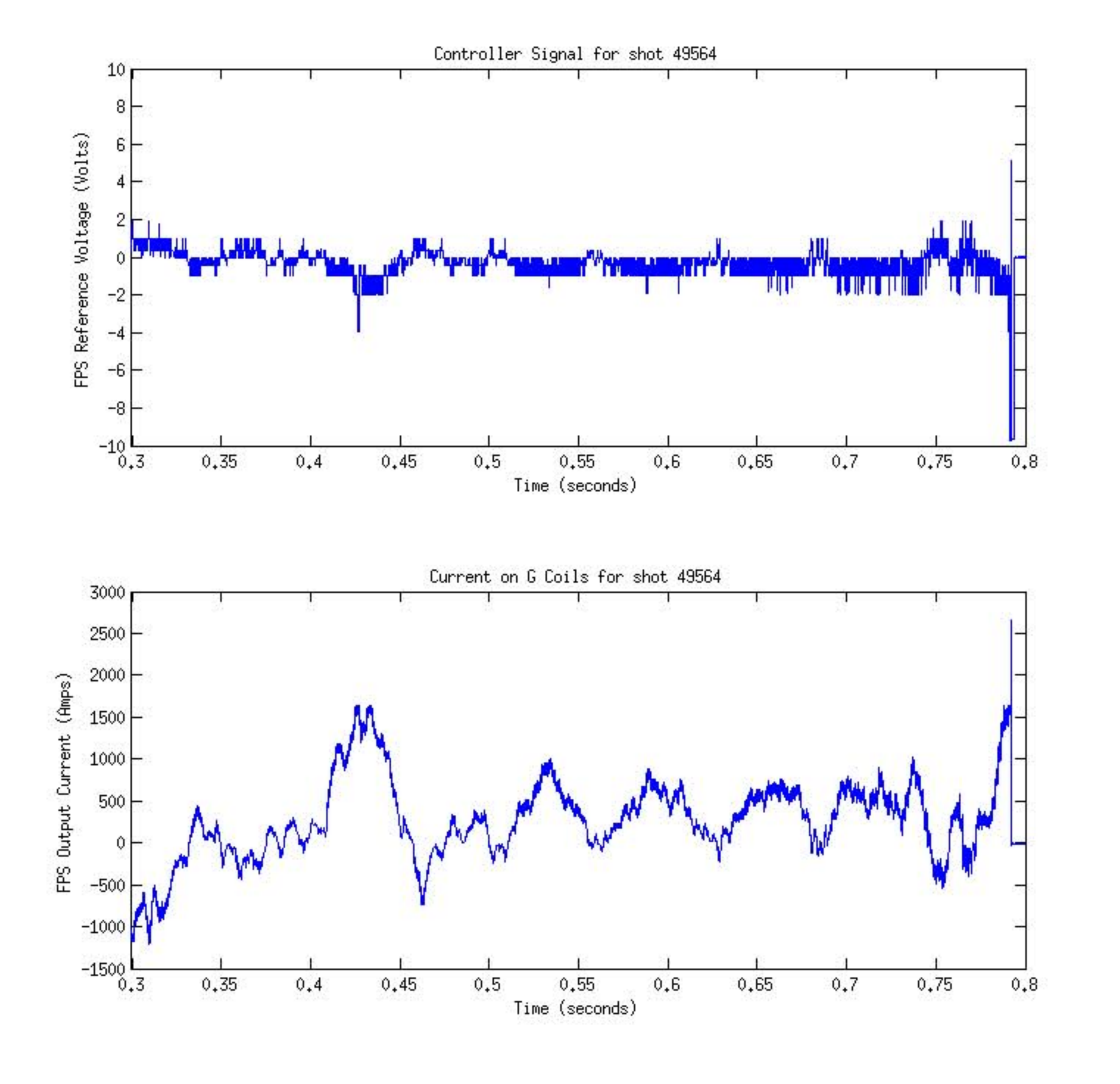}
	\caption{Control signal and coil current for discharge 49564 using the new bang-bang controller}
	\label{fig:FPSControlandCurrents49564}
\end{figure}

\begin{figure}
	\centering
	\includegraphics[scale=0.38]{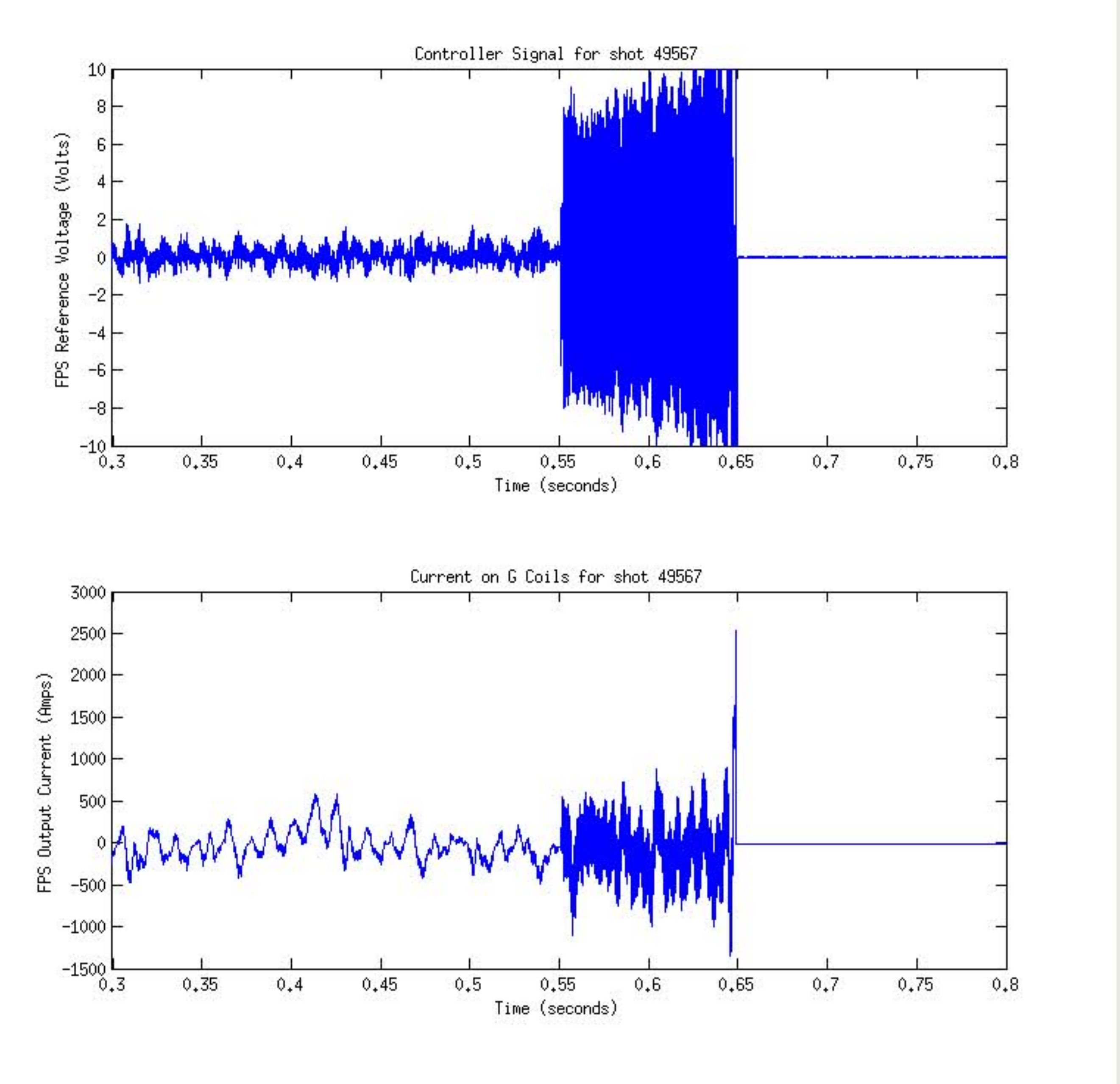}
	\caption{Control signal and coil current for reference discharge 49567 using the standard controller in the same plasma conditions as discharge 49564}
	\label{fig:FPSControlandCurrents49567}
\end{figure}

Figures \ref{fig:PlasmaPositionandVelocity49564} and \ref{fig:PlasmaPositionandVelocity49567} also show a smaller deviation for the plasma position and velocity during the discharge. Figures \ref{fig:FPSControlandCurrents49564} and \ref{fig:FPSControlandCurrents49567} depict a better use of the coil currents. The plasma position and velocity are more stable during the complete discharge without the continuous fast up-down movement that can be seen using the PID controller.

The vertical stabilization controller was implemented and tested using one of the hardware modules with parallel digital signal processing capabilities of the Advanced Plasma Control System \cite{Cruz2008}. For further testing of the controller it is envisaged the use of an ELM detector \cite{ELMS} capable of signaling the error and unavailability of plasma position observer. It is also planned the controller implementation in a newer control hardware based on FPGA \cite{Bao} to study and compare the performance of both systems.

\end{document}